\documentclass[aps,preprint]{revtex4}
\usepackage[T1]{fontenc}
\usepackage[utf8]{inputenc}
\usepackage{array}
\usepackage{multirow}
\usepackage{amsmath}
\usepackage{amssymb}
\usepackage{graphicx}

\makeatletter

\providecommand{\tabularnewline}{\\}

\@ifundefined{textcolor}{}
{%
 \definecolor{BLACK}{gray}{0}
 \definecolor{WHITE}{gray}{1}
 \definecolor{RED}{rgb}{1,0,0}
 \definecolor{GREEN}{rgb}{0,1,0}
 \definecolor{BLUE}{rgb}{0,0,1}
 \definecolor{CYAN}{cmyk}{1,0,0,0}
 \definecolor{MAGENTA}{cmyk}{0,1,0,0}
 \definecolor{YELLOW}{cmyk}{0,0,1,0}
}

\usepackage{amsfonts}
\usepackage{array}
\usepackage{multirow}
\usepackage{latexsym}
\usepackage{epsfig}
\usepackage{float}
\usepackage{dsfont}

\providecommand{\tabularnewline}{\\}

\newcommand{\be}{\begin{equation}}
\newcommand{\ee}{\end{equation}}

\RequirePackage{doi}

\usepackage{caption}
\usepackage{ragged2e}
\justifying\let\raggedright\justifying
\allowdisplaybreaks

\@ifundefined{showcaptionsetup}{}{%
 \PassOptionsToPackage{caption=false}{subfig}}
\usepackage{subfig}
\makeatother

\begin{document}
\title{Strong Cosmic Censorship for a Scalar Field in a Logarithmic-de Sitter
Black Hole}
\author{Yiqian Chen}
\email{chenyiqian@stu.scu.edu.cn}

\author{Qingyu Gan}
\email{gqy@stu.scu.edu.cn}

\author{Guangzhou Guo}
\email{guangzhouguo@stu.scu.edu.cn}

\affiliation{Center for Theoretical Physics, College of Physics, Sichuan University,
Chengdu, 610064, PR China}
\begin{abstract}
It has been shown that the quasinormal modes of perturebated fields
can be used to investigate the validity of strong cosmic censorship
(SCC). Relevant issues for Reissner-Nordstrom-de Sitter (RN-dS) black
holes and Born-Infeld-de Sitter (BI-dS) black holes have been discussed.
In this paper, we investigate SCC in an asymptotic RN-dS black hole
with logarithmic nonlinear electromagnetic field perturbed by massless
scalar fields. It has been argued that SCC can be violated in a near-extremal
RN-dS black hole. However, we find that the NLED effect can rescue
SCC for a near-extremal logarithmic-de Sitter black hole. Compared
with Born-Infeld model, we find that the NLED effect have similar
behavior.
\end{abstract}
\maketitle
\tableofcontents{}

\bigskip{}

%\affiliation{Center for Theoretical Physics, College of Physical Science and Technology,
%Sichuan University, Chengdu, 610064, PR China}

%\affiliation{Center for Theoretical Physics, College of Physical Science and Technology,
%Sichuan University, Chengdu, 610064, PR China}

\section{Introduction}

The Strong Cosmic Censorship (SCC) was proposed by Penrose to maintain
the predictability of general relativity. As we know, a spacetime
singularity can be formed by the gravitational collapse. Singularities
can be classified as space-like singularities, light-like singularities
and time-like singularities. For a spacetime with a time-like singularity,
general relativity will lose its predictability because some regions
in the space time can be influenced by the uncertain data on the singularity.
To solve this problem, SCC asserts that, starting with some physically
relevant initial data for Einstein's equation, the dynamics of physical
systems will always produce globally hyperbolic spacetime \citep{Penrose:1969pc,Penrose:1964wq,Hawking:1969sw}.
In other words, a black hole formed by gravitational collapse or other
physically acceptable dynamical procedures can only has space-like
or light-like singularities, while time-like singularities are forbidden.
However, some solutions of Einstein's equation possess time-like singularity,
such as Reissner-Nordstrom black holes and Kerr-Newman black holes,
which have Cauchy horizons. To fit with SCC, it is required that the
perturbation for any fields at the Cauchy horizon is inextendible
when physical initial data is perturbed. Therefore there is another
statement of SCC: generally speaking, the maximal Cauchy development
is inextendible.

However, the extendibility of the Cauchy horizon has some subtleties.
For this reason, people proposed several formulated versions of SCC.
If the perturbation (about the metric) arising from smooth initial
data is $C^{r}$ non-differentiable at the Cauchy horizon, it's called
$C^{r}$ formulation of SCC \citep{Luk:2017jxq,Luk:2017ofx}. For
example, the $C^{0}$ formulation demands that the perturbed metric
to be noncontinuous at the Cauchy horizon. It indeed satisfies the
requirement that the maximal Cauchy development is inextendible, yet
has been proved to be wrong. There are also lots of discussions about
the $C^{2}$ formulation which corresponds to the divergence of the
curvature. Since the equations of motion are of the second order,
it's reasonable that the curvature is required to be divergent at
the Cauchy horizon. However, the $C^{2}$ formulation is still not
appropriate. A macroscopic observer is able to cross the Cauchy horizon
safely without being destroyed by a divergent curvature, therefore
the $C^{2}$ formulation needs to be strengthened. With weak solutions
showing many important physical applications, it becomes more reasonable
to consider the weak solutions of the equations of motion. This idea
leads to the Christodoulou’s formulation of SCC \citep{Christodoulou:2008nj},
which will be adopted in the following discussion.

For simplicity, we put a test particle (field) into the spacetime
without considering the back-reaction. For a linear massless scalar
field, if SCC is implied, the scalar field perturbation will not belong
to the Sobolev space $H_{loc}^{1}$ at the Cauchy horizon. Such fields
have infinite energy at the Cauchy horizon. It has been proven that
the Christodoulou’s formulation is appropriate in the case of RN black
holes and Kerr black holes. Generally speaking, when the scalar field
propagates to the Cauchy horizon, it will experience an power-law
decay \citep{Price:1971fb,Dafermos:2014cua,Angelopoulos:2016wcv},
at the same time, there is an exponential blue-shift effect \citep{Chambers:1997ef,Dafermos:2003wr,Poisson:1990eh,Ori:1991zz,Hod:1998gy,Brady:1995ni}.
Ultimately, the dominant blue-shift effect would make the Cauchy horizon
become singular \citep{Dafermos:2003wr,Dafermos:2012np}, therefore,
SCC is respected for RN black holes and Kerr black holes. The above
conclusion is based on the case that the cosmological constant $\Lambda$
equals to $0$. If we consider the spacetime with positive cosmological
constant, situations will become quite different. The decay of the
scalar field will be exponential rather than power-law near the Cauchy
horizon \citep{2007arXiv0706.0350B,Dyatlov:2013hba,Dyatlov:2011jd,Hintz:2016gwb,Hintz:2016jak,Berti:2009kk,Konoplya:2011qq,Brady:1996za}.
So the validity of SCC depends on the competition between the exponential
decay and blue-shift effect, which can be characterized by \citep{Costa:2014aia,Costa:2014yha,Costa:2014zha,Hintz:2015jkj,Kehle:2018upl,Cardoso:2017soq}
\begin{equation}
\beta=\frac{\alpha}{\kappa_{-}},\label{eq:beta}
\end{equation}
where $\kappa_{-}$ is the surface gravity of Cauchy horizon, and
$\alpha$ is defined as $\inf_{ln}\left\{ -\mathrm{Im}\left(\omega_{ln}\right)\right\} $,
where the $\omega_{ln}$ is the quasinormal frequencies of the quasinormal
modes (QNMs). (We will give more discussion at Section \ref{sec:Quasinormal-Mode}.)
The criterion is whether $\beta>\frac{1}{2}$ or not; if it's true,
then Christodoulou’s formulation of SCC is violated. It has been shown
that SCC is respected in a Kerr-de Sitter black hole but violated
in a near-extremal RN-dS black hole by a scalar field \citep{Rahman:2018oso,Dias:2018ynt,Cardoso:2017soq}.
Violation of SCC in RN-dS black hole for various other fields, such
as charged scalar field and Dirac field, are also discussed \citep{Mo:2018nnu,Dias:2018ufh,Cardoso:2018nvb,Ge:2018vjq,Destounis:2018qnb}.
When considering the coupled linearized electromagnetic and gravitational
perturbation, the violation of SCC will be severer in a RN-dS black
hole \citep{Dias:2018etb}. What's more, it has been shown that nonlinear
perturbations are not able to prevent SCC from being violated \citep{Luna:2018jfk}.

As we know, the RN metric is a solution of Einstein-Maxwell gravity,
which has the infinite self-energy for charged point-like particles.
Moreover, a point charge can not only lead to the electromagnetic
singularity, but also the spacetime singularity through the gravitational
field equations. Before renormalization, a classical approach has
been proposed to solve this problem, namely the nonlinear electrodynamics
(NLED). This approach was later generalized and applied to many other
problems, like the limiting curvature hypothesis in cosmological theories
and the vacuum polarization effect \citep{Mukhanov:1991zn,Brandenberger:1993ef,Heisenberg:1935qt}.
NLED was first introduced in the 1930s by Born and Infeld (BI). In
addition to the above advantages, their nonlinear electrodynamics
can also serve as a low energy effective limit of the superstring
theory and play roles in the AdS/CFT correspondence \citep{Fradkin:1985qd}.
Various NLED modes have been proposed and investigated for different
purposes, such as exponential electrodynamics and logarithmic electrodynamics
\citep{Wang:2018hwg,Wang:2018xdz,Wang:2019dzl,Wang:2019kxp}. In this
paper, we investigate the logarithmic electrodynamics, which can also
remove the infinite self-energy. Although it does not have a background
in superstring theory, it is still a good toy model to study various
interesting subjects. When expanded to the second order of the NLED
parameter, the action of logarithmic electrodynamics is consistent
with that of the BI electrodynamics\citep{Soleng:1995kn}.

Two of us have discussed the validity of SCC in a BI-dS black hole
in \citep{Gan:2019jac}. In order to explore the similarities and
differences among different NLED effects, we further investigate SCC
in a logarithmic-de Sitter black hole. Our numerical results show
that NLED effect of the BI electrodynamics and the logarithmic electrodynamics
are similar in essential while differ in minor points. The most important
conclusion is that the NLED effect can restore SCC in the near-extremal
regime, which is violated in the RN-dS black hole.

In Section \ref{sec:Logarithmic-dS-Black-Hole}, we review the logarithmic
black hole solution. In Section \ref{sec:Quasinormal-Mode}, we introduce
the quasinormal modes method. In Section \ref{sec:Numerical-Results},
we present our numerical results and give some discussions. It is
worth mentioning that our analysis is based on the Christodoulou’s
formulation of SCC, and we set $16\pi G=c=1$ throughout this paper.

\section{Logarithmic-dS Black Hole\label{sec:Logarithmic-dS-Black-Hole}}

In this section, we review the black hole solution with logarithmic
electromagnetic field. Then we find the parameter regions which allow
three horizons so that the Cauchy horizon exists.

First Let's begin with the action with logarithmic electromagnetic
field 
\begin{equation}
S=\int d^{4}x\sqrt{-g}\left[\mathcal{R}-2\Lambda+L(\mathcal{F})\right],\label{eq:action}
\end{equation}
\begin{equation}
L(\mathcal{F})=-8b^{2}\ln\left(1+\frac{\mathcal{F}}{8b^{2}}\right),
\end{equation}
where $\mathcal{R}$ is the Ricci scalar, $\mathcal{F}=F_{\mu\nu}F^{\mu\nu}$,
$F_{\mu\nu}=\partial_{\mu}A_{\nu}-\partial_{\nu}A_{\mu}$ is the electromagnetic
field, $A_{\mu}$ is the corresponding electromagnetic potential,
and $b$ is the NLED parameter. NLED effect will magnify as the parameter
$b$ becomes small. On the contrary, NLED effect will reduce when
the parameter $b$ increases. The logarithmic electrodynamics recovers
Maxwell electrodynamics in the limit of $b\rightarrow\infty$.

Varying the action (\ref{eq:action}), it's not hard to get the equations
of motion 
\begin{equation}
R_{\mu\nu}-\frac{1}{2}g_{\mu\nu}(R-2\Lambda)=\frac{1}{2}g_{\mu\nu}L(\mathcal{F})-2F_{\mu\sigma}F_{\nu}^{\ \sigma}L_{\mathcal{F}},\label{eq:Eeq}
\end{equation}
\begin{equation}
\partial_{\mu}(\sqrt{-g}L_{\mathcal{F}}F^{\mu\nu})=0,\label{eq:EOM}
\end{equation}
where $R_{\mu\nu}$ is the Ricci tensor, and $L_{\mathcal{F}}\equiv\frac{dL(\mathcal{F})}{d\mathcal{F}}$.
To demand a static spherically symmetric black hole solution, the
metric of the logarithmic-dS black hole was obtained in \citep{Sheykhi:2015hdc}

\[
ds^{2}=-f\left(r\right)dt^{2}+\frac{dr^{2}}{f\left(r\right)}+r^{2}(d\theta^{2}+\sin^{2}\theta d\varphi^{2}),
\]
\begin{eqnarray}
f\left(r\right) & = & 1-\frac{\left(\Lambda-4b^{2}\right)r^{2}}{3}-\frac{M}{8\pi r}+\frac{8b^{2}}{9}r^{2}\left(1-\thinspace_{2}\mathcal{F}_{1}\left[-\frac{1}{2},-\frac{3}{4};\frac{1}{4};-\frac{Q^{2}}{b^{2}r^{4}}\right]\right)\label{eq:metric}\\
 &  & -\frac{4b^{2}}{3}r^{2}\left(\sqrt{\frac{Q^{2}}{b^{2}r^{4}}+1}-\log\left(\frac{Q^{2}}{2b^{2}r^{4}}\right)+\log\left(\sqrt{\frac{Q^{2}}{b^{2}r^{4}}+1}-1\right)\right).\nonumber 
\end{eqnarray}
After fixing gauge, the logarithmic electromagnetic potential is 
\begin{equation}
\mathbf{A}=A_{t}dt=\frac{2Q}{3r}\left(\frac{1}{1+\sqrt{1+\frac{Q^{2}}{r^{4}b^{2}}}}-2\thinspace_{2}\mathcal{F}_{1}\left[\frac{1}{4},\frac{1}{2};\frac{5}{4};-\frac{Q^{2}}{b^{2}r^{4}}\right]\right)dt,\label{eq:electromagnetic field}
\end{equation}
here $\ _{2}\mathcal{F}_{1}$ is the hypergeometric function, $M$
and $Q$ are the mass and electric charge of the logarithmic-dS black
hole, respectively. In the limit of $b\rightarrow\infty$, the metric
(\ref{eq:metric}) and the electromagnetic potential (\ref{eq:electromagnetic field})
recover the RN-dS black hole solution as expected.

To investigate SCC, we need to calculate QNMs at the Cauchy horizon,
hence we only focus on the logarithmic-dS black holes which possess
three horizons. It means that we need to find out the allowed parameter
region where $f(r)=0$ has three positive solutions, which correspond
to the positions of the Cauchy horizon $r_{-}$, the event horizon
$r_{+}$ and the cosmological horizon $r_{c}$, respectively. Because
of the complexity of $f(r)$, we can only find the allowed region
by numerical method and a bit analysis. First of all, through numerical
simulation, we find that an appropriate $f(r)$ always has two extreme
points $r_{min}$ and $r_{max}$, which doesn't coincide with zero
points, namely, $f(r_{min})<0$ and $f(r_{max})>0$. It is noteworthy
that there are extremal black hole solutions with $r_{-}=r_{+}$ and
$f(r_{min})=0$, in which situation we denote the charge of black
hole as $Q_{ext}$. Similarly, there are solutions with $r_{+}=r_{c}$
and $f(r_{max})=0$, known as the Nariai black holes \citep{Ginsparg:1982rs,Bousso:1996au}.
These solutions form the bounds of the allowed region in the parameter
space. Note that $f(r)\rightarrow-\infty$ in the limit $r\rightarrow+\infty$,
therefore, $f(r)$ is supposed to be positive in the limit $r\rightarrow0$.
We expand $f(r)$ near $r=0$ as follows 
\begin{equation}
f(r)=\left(-\frac{M}{8\pi}+\frac{4bQ\Gamma\left(\frac{1}{4}\right)^{2}\sqrt{\frac{Q}{b}}}{9\sqrt{\pi}}\right)r^{-1}+\left(-\frac{4bQ}{3}+\frac{2bQ\Gamma\left(-\frac{1}{4}\right)}{3\Gamma\left(\frac{3}{4}\right)}+1\right)+O\left(r\right),\label{eq: fr expand}
\end{equation}
where $\Gamma$ is the Gamma function. Hence we get a relation 
\begin{equation}
-\frac{M}{8\pi}+\frac{4bQ\Gamma^{2}\left(\frac{1}{4}\right)\sqrt{\frac{Q}{b}}}{9\sqrt{\pi}}>0.\label{eq:constraint1}
\end{equation}
For simplicity, we take $M=16\pi$ from now on without losing generality.
The constraint (\ref{eq:constraint1}) between $Q$ and $b$ becomes
\begin{equation}
bQ^{3}>k,\ \ k=\frac{81\pi}{4\Gamma^{2}\left(\frac{1}{4}\right)}.\label{eq:c2}
\end{equation}

\begin{figure}[t]
\begin{centering}
\includegraphics{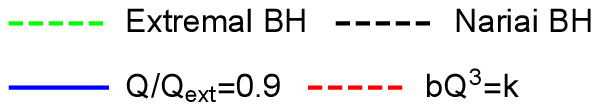}
\par\end{centering}
\ 
\begin{centering}
\includegraphics[scale=0.6]{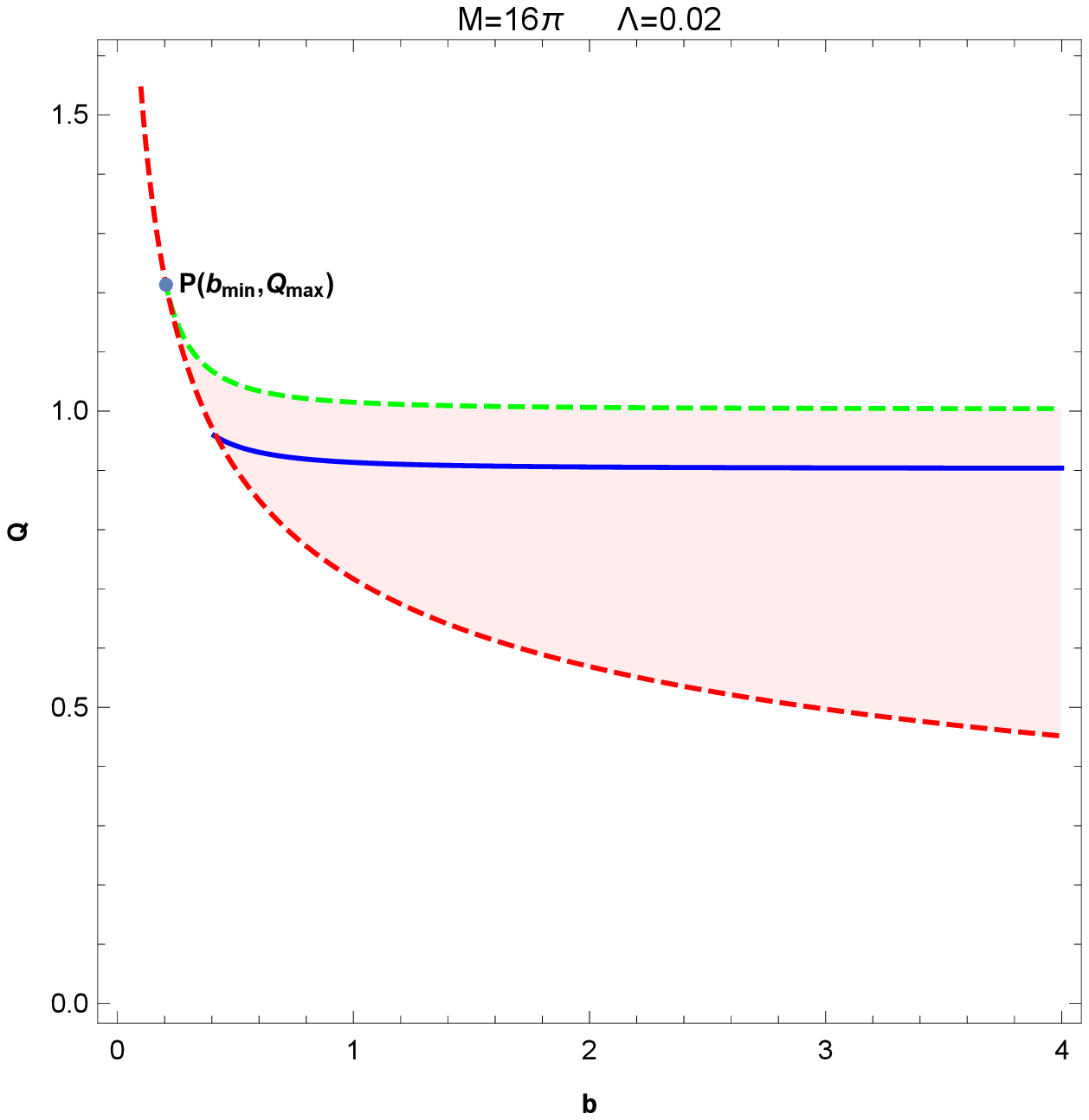}\  \includegraphics[scale=0.6]{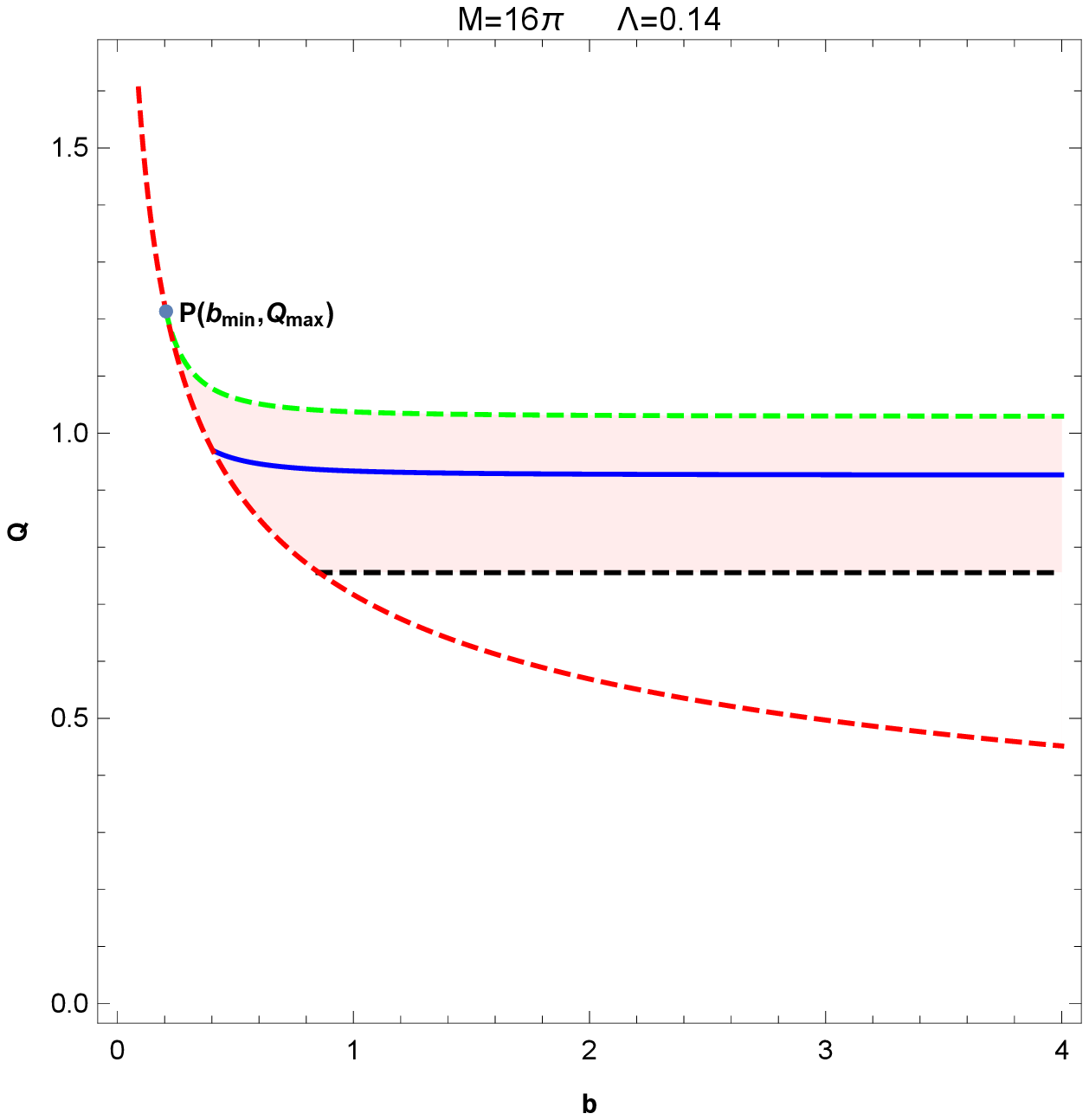}
\par\end{centering}
\caption{Parameter space for logarithmic-ds black holes with $M=16\pi$. Left:
$\Lambda=0.02$. Right: $\Lambda=0.14$. The pink region allows for
black holes with three horizons. The dashed green lines represent
the extremal black holes with $r_{-}=r_{+}$; the dashed black line
represents the Nariai black hole with $r_{+}=r_{c}$; and the dashed
red lines represent the constraint $bQ^{3}=k$. We can see the dashed
red line and the dashed green line intersect at the point $P(b_{min},Q_{max})$.
Here, $b_{min}\thickapprox0.206008$ and $Q_{max}\thickapprox1.213543$
are calculated by numerical method which doesn't depend on $\Lambda$
(an analytic expression will be given in the Section \ref{sec:Numerical-Results}).
The blue line represent near-extremal black holes with charge $Q=0.9Q_{ext}$,
where $Q_{ext}$ is the charge of a extremal black hole with the same
$b$.}
\label{region}
\end{figure}

Based on the above analysis, we plot the allowed region with three
horizons in FIG.\ref{region}. As shown in the figure, the case $\Lambda=0.02$
is different from the case $\Lambda=0.14$. The parameter space of
$\Lambda=0.14$ has Nariai black holes, while the parameter space
of $\Lambda=0.02$ does not. Actually, this is not unique to the logarithmic-dS
black holes. RN-dS black holes also have the similar conclusion. However,
we will only be interested in the region near $P$, then the difference
between various $\Lambda$ is not significant. Near the point $P$,
the allowed region is bounded by the upper bound (the green dashed
line) and lower bound (the red dashed line).

\section{Quasinormal Mode\label{sec:Quasinormal-Mode}}

To investigate SCC, we need to take a test field as a probe to perturb
the logarithmic-dS black hole. In this section, Let's consider a scalar
field perturbation with mass $\mu$ and charge $q$. The equation
of motion of the field is the Klein-Gordon equation in a curved spacetime
\begin{equation}
\left(\mathbf{D}^{2}-\mu^{2}\right)\Phi=0,\label{eq:KG eq.}
\end{equation}
where $\mathbf{D}$ is the covariant derivative $\mathbf{D}=\mathbf{\mathbf{\mathbf{\nabla}}}-iq\mathbf{A}$.
We make a transformation $v\equiv t+r_{*}$, where $r_{*}$ is the
tortoise coordinate defined as 
\begin{equation}
r_{*}=\int\frac{dr}{f(r)}.\label{eq:rstar}
\end{equation}
Then we define the Eddington-Finkelstein ingoing coordinates $(v,r,\theta,\varphi)$
like in \citep{Dias:2018etb}, so that we can discuss the properties
of horizons more straightforward. Moreover, the gauge fixed logarithmic
electromagnetic potential is 
\begin{equation}
\mathbf{A}=A_{v}dv=\frac{2Q}{3r}\left(\frac{1}{1+\sqrt{1+\frac{Q^{2}}{r^{4}b^{2}}}}-2\thinspace_{2}\mathcal{F}_{1}\left[\frac{1}{4},\frac{1}{2};\frac{5}{4};-\frac{Q^{2}}{b^{2}r^{4}}\right]\right)dv.\label{eq:gauge field}
\end{equation}
Since we demand the logarithmic-dS black hole solution to be static
and spherically symmetric, we can separate the field solution as the
standard ansatz

\begin{equation}
\Phi=e^{-i\omega v}Y_{lm}\left(\theta,\phi\right)\psi_{\omega l}\left(r\right),\label{eq:=00005CPhi}
\end{equation}
where $Y_{lm}\left(\theta,\phi\right)$ is the spherical harmonic
function, and $\psi_{\omega l}\left(r\right)$ is the radial function.
Plugging the ansatz (\ref{eq:=00005CPhi}) into the equation of motion
(\ref{eq:KG eq.}), we get the radial equation
\begin{eqnarray}
0 & = & \left[r^{2}f\partial_{r}^{2}+\left(r^{2}f^{\prime}+2rf-2iqA_{v}r^{2}-2i\omega r^{2}\right)\partial_{r}\right.\nonumber \\
 &  & \left.-2i\omega r-2iqrA_{v}-iqr^{2}\partial_{r}A_{v}-l\left(l+1\right)-\mu^{2}r^{2}\right]\psi_{\omega l}\left(r\right),\label{eq:radial-equation}
\end{eqnarray}
where $f^{\prime}$ denotes $df(r)/dr$. For the convenience of numerical
calculation, we define a new coordinate $x\equiv(r-r_{+})/(r_{c}-r_{+})$,
such that the event horizon and cosmological horizon are located at
$x=0,1$, respectively. Through the Frobenius method, we can impose
the boundary solutions near the event horizon and the cosmological
horizon. Near the event horizon, the ingoing and outgoing boundary
solutions are
\begin{equation}
\psi_{\omega l}^{\textrm{ingoing}}\sim\textrm{const},\quad\psi_{\omega l}^{\textrm{outgoing}}\sim x^{i\frac{\omega+q\left.A_{v}\right|_{r=r_{+}}}{\kappa_{+}}}.
\end{equation}
Similarly, near the cosmological horizon, the ingoing and outgoing
boundary solutions are
\begin{equation}
\psi_{\omega l}^{\textrm{ingoing}}\sim\textrm{const},\quad\psi_{\omega l}^{\textrm{outgoing}}\sim(1-x)^{-i\frac{\omega+q\left.A_{v}\right|_{r=r_{c}}}{\kappa_{c}}}.
\end{equation}
where $\kappa_{h}\equiv\left|f^{\prime}\left(r_{h}\right)\right|/2$
denotes the surface gravity at each horizon with $h\in\{+,-,c\}$.
These conditions select a discrete set of frequencies $\omega_{ln}$,
namely the quasinormal frequencies, where $l$ is the angular number,
and $n$ is the mode number. The corresponding modes are called quasinormal
modes (QNMs). It is noteworthy that there is a zero mode when $l=0$,
which should be ignored\citep{Cardoso:2017soq}. In fact, there are
a variety of methods to compute the QNMs. Our numerical results in
section \ref{sec:Numerical-Results} are based on the Chebyshev collocation
scheme, and obtained mostly by the Mathematica package provided in
\citep{Jansen:2017oag,Jansen,url-centra.tecnico.}. In order to fit
the numerical scheme, we redefine the field $\psi_{\omega l}$ as

\begin{equation}
\psi_{\omega l}=\frac{1}{x}\left(1-x\right)^{-i\frac{\omega+q\left.A_{v}\right|_{r=r_{c}}}{\kappa_{c}}}\phi_{\omega l},\label{eq:redefined-phi}
\end{equation}
so that the redefined field $\phi_{\omega l}$ become regular at both
the event horizon and the cosmological horizon.

\section{Numerical Results\label{sec:Numerical-Results}}

\begin{table}
\begin{centering}
\begin{tabular}{|c|c|c|c|c|c|c|}
\hline 
$\Lambda$ & $b$ & $Q/Q_{ext}$ & $q$ & $l=0$ & $l=1$ & $l=10$\tabularnewline
\hline 
\hline 
\multirow{8}{*}{0.02} & \multirow{4}{*}{0.5} & \multirow{2}{*}{0.991} & 0 & 0 & $-0.472594i$ & $\pm14.968405-0.467179i$\tabularnewline
\cline{4-7} \cline{5-7} \cline{6-7} \cline{7-7} 
 &  &  & 0.1 & $0.059183+0.003005i$ & $0.033367-0.471978i$ & $15.286357-0.467124i$\tabularnewline
\cline{3-7} \cline{4-7} \cline{5-7} \cline{6-7} \cline{7-7} 
 &  & \multirow{2}{*}{0.996} & 0 & 0 & $\pm3.605333-0.789770i$ & $\pm25.275405-0.770342i$\tabularnewline
\cline{4-7} \cline{5-7} \cline{6-7} \cline{7-7} 
 &  &  & 0.1 & $0.099702+0.005331i$ & $4.165818-0.779759i$ & $25.818396-0.769624i$\tabularnewline
\cline{2-7} \cline{3-7} \cline{4-7} \cline{5-7} \cline{6-7} \cline{7-7} 
 & \multirow{4}{*}{10000} & \multirow{2}{*}{0.991} & 0 & 0 & $-0.475688i$ & $\pm14.365381-0.491756i$\tabularnewline
\cline{4-7} \cline{5-7} \cline{6-7} \cline{7-7} 
 &  &  & 0.1 & $0.057773+0.002229i$ & $0.032203-0.475118i$ & $-14.080016-0.491441i$\tabularnewline
\cline{3-7} \cline{4-7} \cline{5-7} \cline{6-7} \cline{7-7} 
 &  & \multirow{2}{*}{0.996} & 0 & 0 & $-0.789379i$ & $\pm23.969407-0.808962i$\tabularnewline
\cline{4-7} \cline{5-7} \cline{6-7} \cline{7-7} 
 &  &  & 0.1 & $0.096356+0.003870i$ & $0.053708-0.788423i$ & $-23.488922-0.808825i$\tabularnewline
\hline 
\multirow{8}{*}{0.06} & \multirow{4}{*}{0.5} & \multirow{2}{*}{0.991} & 0 & 0 & $\pm2.021008-0.458730i$ & $\pm14.396115-0.441376i$\tabularnewline
\cline{4-7} \cline{5-7} \cline{6-7} \cline{7-7} 
 &  &  & 0.1 & $0.128077+0.003969i$ & $2.384318-0.452275i$ & $14.744827-0.441085i$\tabularnewline
\cline{3-7} \cline{4-7} \cline{5-7} \cline{6-7} \cline{7-7} 
 &  & \multirow{2}{*}{0.996} & 0 & 0 & $\pm3.431580-0.759489i$ & $\pm24.467637-0.730447i$\tabularnewline
\cline{4-7} \cline{5-7} \cline{6-7} \cline{7-7} 
 &  &  & 0.1 & $0.216661+0.007511i$ & $4.054003-0.743932i$ & $25.066143-0.729315i$\tabularnewline
\cline{2-7} \cline{3-7} \cline{4-7} \cline{5-7} \cline{6-7} \cline{7-7} 
 & \multirow{4}{*}{10000} & \multirow{2}{*}{0.991} & 0 & 0 & $\pm1.930716-0.481345i$ & $\pm13.798347-0.462716i$\tabularnewline
\cline{4-7} \cline{5-7} \cline{6-7} \cline{7-7} 
 &  &  & 0.1 & $0.127461+0.001895i$ & $2.265562-0.474726i$ & $14.119498-0.462581i$\tabularnewline
\cline{3-7} \cline{4-7} \cline{5-7} \cline{6-7} \cline{7-7} 
 &  & \multirow{2}{*}{0.996} & 0 & 0 & $\pm3.242616-0.795833i$ & $\pm23.189760-0.764924i$\tabularnewline
\cline{4-7} \cline{5-7} \cline{6-7} \cline{7-7} 
 &  &  & 0.1 & $0.213619+0.003591i$ & $3.808829-0.781460i$ & $23.733891-0.764259i$\tabularnewline
\hline 
\end{tabular}
\par\end{centering}
\caption{The lowest-lying QNMs $\omega/\kappa_{-}$ of different angular numbers
$l$ for various values of $\Lambda$, $b$, $Q/Q_{ext}$ and $q$
for massless scalar perturbation. In the large $b$ limit ($b=10000$),
the numerical results go back to that of the RN-dS black holes \citep{Cardoso:2017soq,Mo:2018nnu}.}

\label{table-QNM}
\end{table}

In this section, we present our numerical results. In the first subsection,
we discuss the validity of SCC with massless neutral scalar perturbations;
in the second subsection, we discuss the validity of SCC with massless
charged scalar perturbations. Since the NLED effect is strong when
$b$ is relatively small, SCC is most likely to be violated in near-extremal
black holes. We are hence more interested in the black holes which
are near the point $P$ in the parameter space.

To verify the reliability of the program, we calculated a series of
lowest-lying QNMs in Table \ref{table-QNM}. Comparing our results
with that of RN-dS black holes \citep{Cardoso:2017soq,Mo:2018nnu},
we find that they are consistent for large $b$ ($b=10000$).

\subsection{Neutral Scalar Field\label{subsec:Neutral-Scalar-Field}}

Since it's impossible to calculate $\omega_{ln}$ for all $l$ and
$n$, a clever method was proposed to seek out the lowest-lying QNMs
for RN-dS black holes \citep{Cardoso:2017soq}. The authors found
three different families that can classify the QNMs: the photon sphere
(PS) modes, with dominant mode at large $l$ ($l=10$ is good enough);
the de Sitter (dS) modes, with dominant mode at $l=1$; and the near-extremal
(NE) modes, with dominant mode at $l=0$. In the process of numerical
calculation, we also find these three distinct families of modes for
a neutral massless scalar field in the logarithmic-dS black hole,
therefore, we are going to discuss the neutral case by these three
families in this subsection.

\begin{figure}
\begin{centering}
\includegraphics{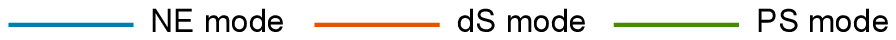}
\par\end{centering}
\begin{centering}
\subfloat[{\small{}The lowest-lying QNMs $-\textrm{Im}(\omega)/\kappa_{-}$of
three families with varying $Q/Q_{ext}$ for various values of $b$
and $\Lambda$. The vertical thin dashed lines indicate that the NE
modes become dominant. On the right side of the thick dashed lines,
we can see that SCC is violated.}]{\begin{centering}
\includegraphics[scale=0.4]{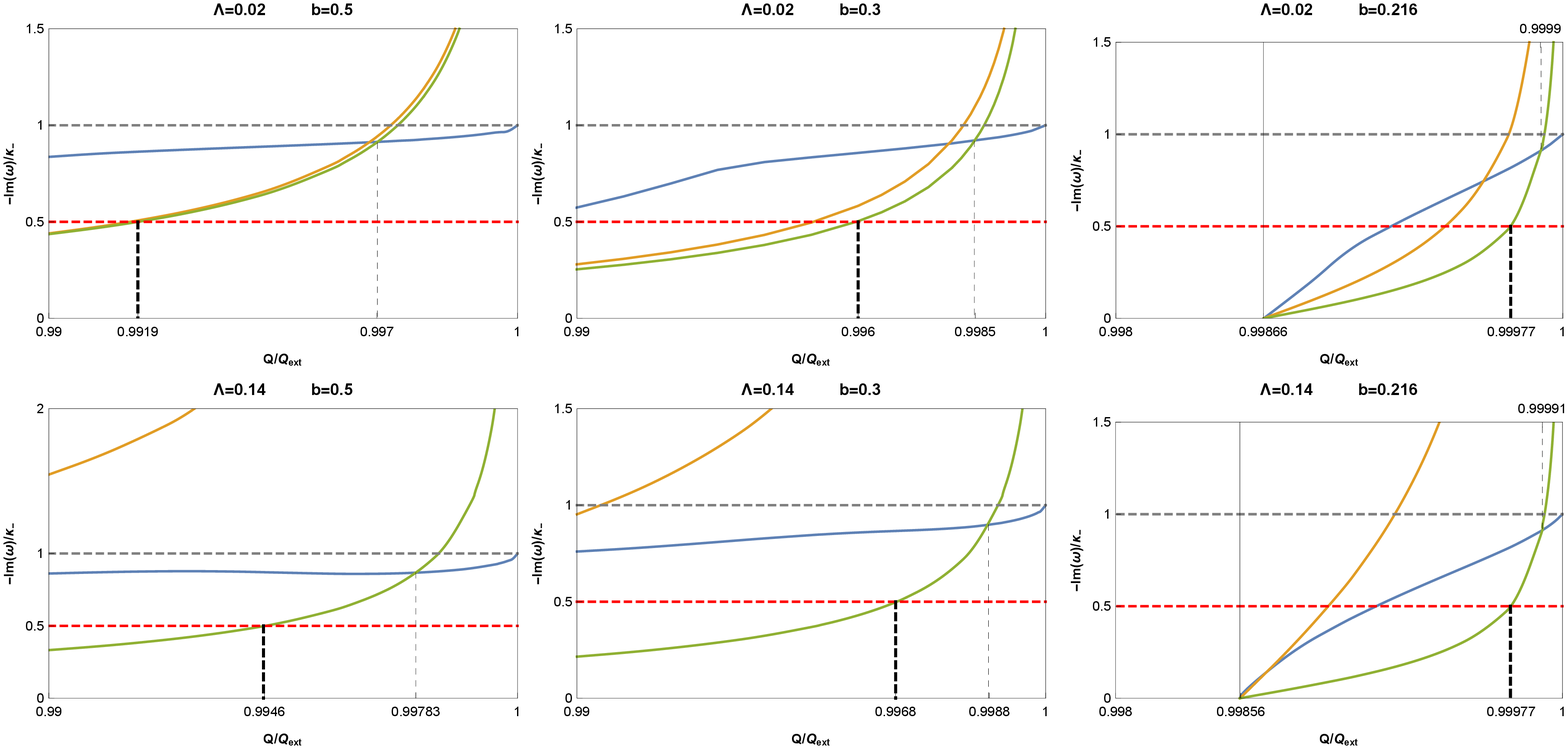}
\par\end{centering}

\label{figure-three-fam-fix-b}}
\par\end{centering}
\begin{centering}
\subfloat[{\small{}The lowest-lying QNMs $-\textrm{Im}(\omega)/\kappa_{-}$
of three families with varying $b$ for various values of $Q/Q_{ext}$
and $\Lambda$. Near the vertical solid thin lines, we can see that
}SCC{\small{} is always }respected{\small{} for a small enough value
of $b$.}]{\begin{centering}
\includegraphics[scale=0.4]{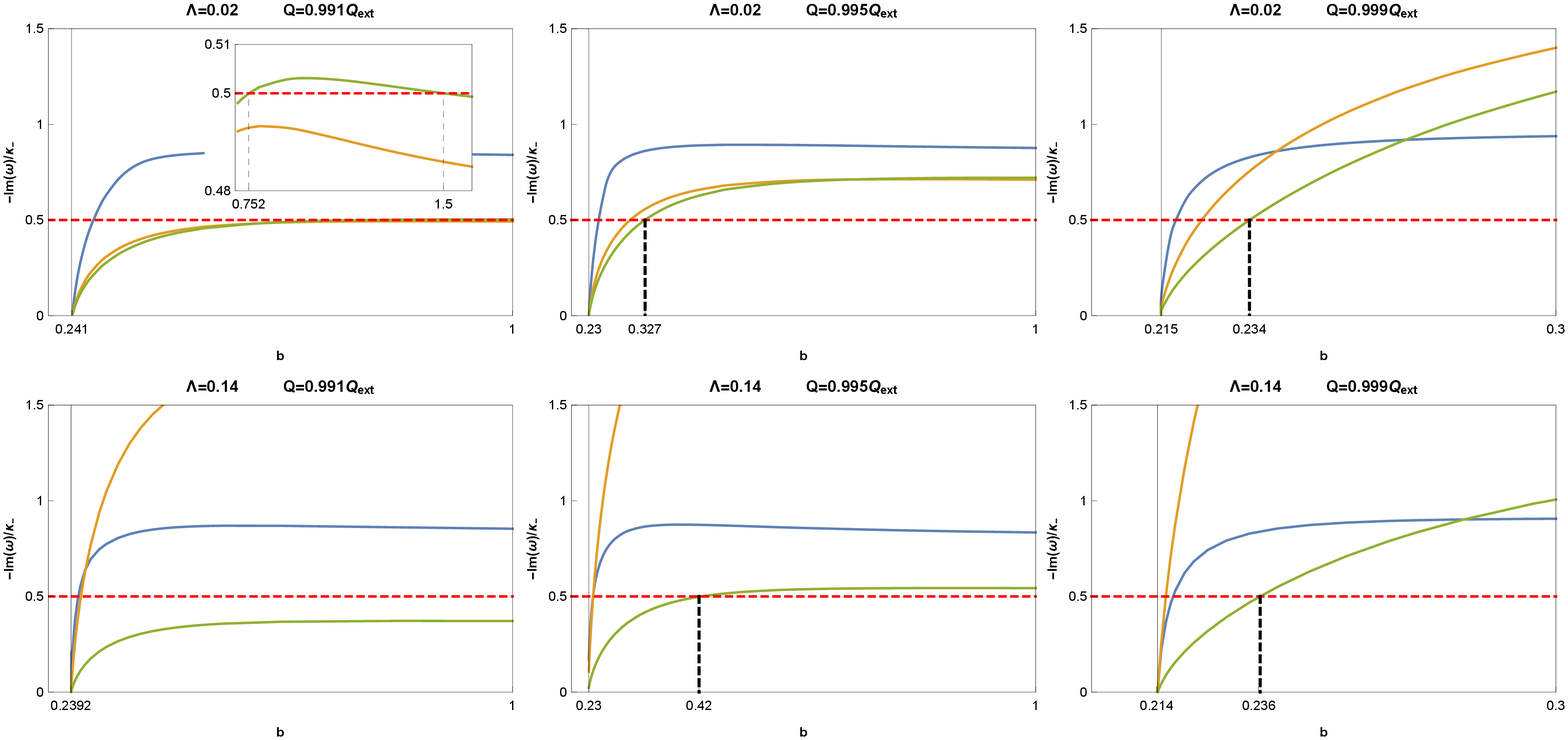}
\par\end{centering}

\label{figure-three-fam-fix-q}}
\par\end{centering}
\caption{The lowest-lying QNMs $-\textrm{Im}(\omega)/\kappa_{-}$ of three
families for a neutral massless scalar field. The vertical solid lines
indicate that the parameters reach the lower bound of allowed region,
where the logarithmic-dS black hole does not have three horizons.
SCC is violated only when the dominant modes of three families are
all above the red dashed line. And the thick black dashed lines indicate
the key points where $\beta\equiv-\textrm{Im}(\omega)/\kappa_{-}=\frac{1}{2}$.}

\label{figure-three-family}
\end{figure}

Before talking about SCC for near-extremal black hole, we first investigate
the behavior near the lower bound given by constraint (\ref{eq:c2}
i.e. the red dashed line in FIG.\ref{region}). We find that the radius
$r_{-}$ goes to zero in the limit of $bQ^{3}\rightarrow k$, so we
can expand the surface gravity of Cauchy horizon around $r_{-}=0$
as follows

\begin{equation}
\kappa_{-}=\left|\frac{1}{2r}\frac{d\left(rf\left(r\right)\right)}{dr}\mid_{r=r_{-}}\right|=\frac{1-4Qb}{2r_{-}}+O(1).
\end{equation}
Therefore as long as $1-4Qb\neq0$, $\kappa_{-}$ will go to infinity
at the lower bound. Indeed, the condition $1-4Qb=0$ and the condition
$bQ^{3}=k$ intersect at the point $P(b_{min},Q_{max})$, where
\[
b_{min}=\frac{4\sqrt{\pi}\Gamma\left(\frac{1}{4}\right)^{2}}{9M},\ \ Q_{max}=\frac{9M}{16\sqrt{\pi}\Gamma\left(\frac{1}{4}\right)^{2}}.
\]
We find $\alpha$ are finite at the lower bound, hence $\beta=0<\frac{1}{2}$.
It means that SCC is always respected for the black holes whose parameters
are close enough to the lower bound (this conclusion is also reflected
in FIG.\ref{figure-three-family} as explained below).

We plot the lowest-lying QNMs $-\textrm{Im}(\omega)/\kappa_{-}$ of
three families in FIG.\ref{figure-three-family}, where the blue line
represents the dominant mode of NE family ($l=0$), the orange line
represent the dominant mode of dS family ($l=1$), the green line
represents the dominant mod of PS family ($l=10$). In FIG.\ref{figure-three-fam-fix-b},
we plot $-\textrm{Im}(\omega)/\kappa_{-}$ against $Q/Q_{ext}$ and
find that the lowest-lying QNMs of PS and dS families go to infinite
while the modes of NE family go to $1$. This behavior not only indicates
that the NE modes become dominant for a sufficiently near-extremal
black hole, but also implies that SCC can always be violated for a
sufficiently near-extremal black hole. Moreover, we find that the
critical $Q/Q_{ext}$ of $\beta=\frac{1}{2}$ increase as $b$ decrease,
which means that SCC is more difficult to be violated as $b$ decreases.
In FIG.\ref{figure-three-fam-fix-q}, we plot $-\textrm{Im}(\omega)/\kappa_{-}$
against $b$ for some fixed $Q/Q_{ext}$. First of all, the most noticeable
feature is that the lowest-lying QNMs of the three families approach
$0$ at $b=b_{Q/Q_{ext}}$, which is consistent with the analysis
in the last paragraph. Here, $b_{Q/Q_{ext}}$ is the minimal $b$
for a fixed charge ratio $Q/Q_{ext}$, namely the intersection of
the line of fixed $Q/Q_{ext}$ and the lower bound as show in FIG.\ref{region}.
In the case of $\Lambda=0.14$, $Q=0.991Q_{ext}$, the lowest-lying
PS mode is lower than the red dashed line for any $b$, therefore
SCC is always respected. Similarly, in the case of $\Lambda=0.02$,
$Q=0.991Q_{ext}$, although the lowest-lying PS mode fluctuates around
$\beta=\frac{1}{2}$, the dominant dS mode save the validity of SCC.
In summary, \textbf{$b$ }and $Q/Q_{ext}$ both have important effects
on the validity of SCC. Specifically, SCC is more likely to be violated
when $Q/Q_{ext}$ goes to $1$, and more likely to be respected when
$b$ goes to $b_{Q/Q_{ext}}$.

\begin{figure}
\begin{centering}
\includegraphics[scale=0.8]{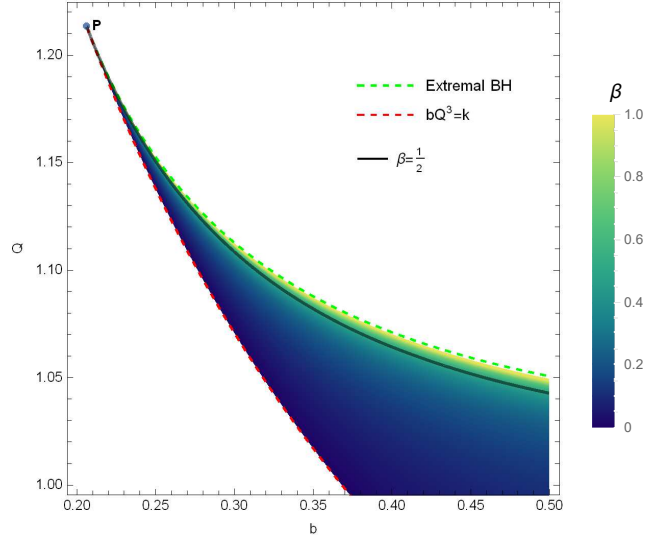}
\par\end{centering}
\caption{$\beta$ for a neutral massless scalar field with $M=16\pi$, and
$\Lambda=0.06$. The green dashed line and the red dashed line correspond
to the extremal black hole ($Q=Q_{ext}$) and lower bound ($bQ^{3}=k$)
of the allowed region, respectively. The solid thick black line represents
$\beta=\frac{1}{2}$, which divides the parameter space into two parts,
where SCC is violated above the black line while respected under it.}
\label{densityplot}
\end{figure}

At the end of this subsection, in order to understand the behavior
of $\beta$ more intuitively, we roughly draw a density plot of $\beta$
near the point $P$ by WKB method in FIG.\ref{densityplot}. In this
figure, we use the PS modes to approximate $\beta$ of non extremal
black holes, where dS modes are similar to PS modes and NE modes are
not dominant. The solid black line, the red dashed line and the green
dashed line represent $\beta=\frac{1}{2}$, $0$, and $1$, respectively.
Note that, SCC is violated between the green line and the black line
($0<\beta<\frac{1}{2}$), while respected between the red line and
the black line ($\frac{1}{2}<\beta<1$).

\subsection{Charged Scalar Field\label{subsec:Charged-Scalar-Field}}

\begin{figure}
\begin{centering}
\includegraphics[scale=0.45]{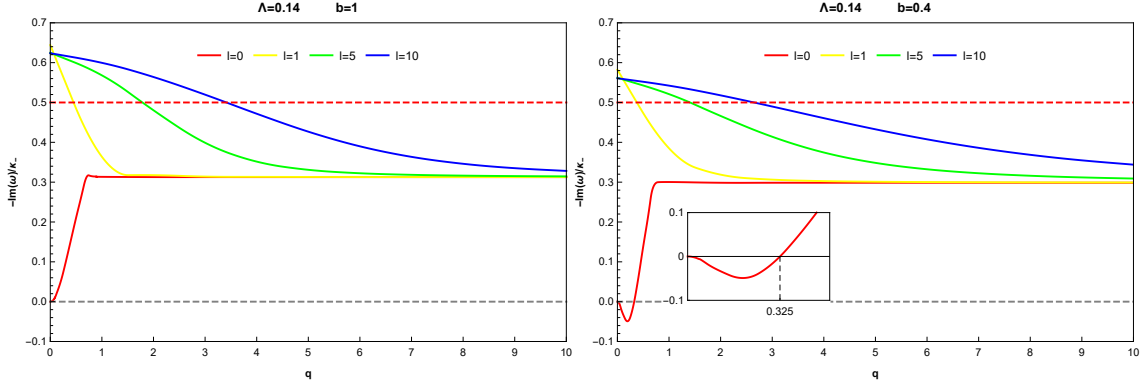}
\par\end{centering}
\caption{The lowest-lying QNMs $-\textrm{Im}(\omega)/\kappa_{-}$ of various
values of $l$ with varying $q$ for $\Lambda=0.14$ and $Q=0.996Q_{ext}$.
Left: $b=1$. Right: $b=0.4$, where the small figure zooms in the
superradiance.}
\label{f4}
\end{figure}

Now we investigate the validity of SCC for a massless charged scalar
field. Unlike the neutral case, we don't use the three families to
classify the QNMs. Since the violation of SCC is more likely to occur
in a near-extremal black hole, we only consider some near-extremal
black holes in the following. In FIG.\ref{f4}, we plot the lowest-lying
QNMs $-\textrm{Im}(\omega)/\kappa_{-}$ of various angular number
$l$ in near-extremal logarithmic-dS black holes. It's easy to find
that the lowest-lying QNMs of $l=0$ (red line) dominate $\beta$
for the charged scalar field in near-extremal black holes. In the
case of $b=0.4$, it is noteworthy that the lowest-lying QNMs can
be negative when scalar charge $q$ is small. Actually, this abnormality
has been found in RN-dS black holes, and was regarded as superradiant
instability \citep{Konoplya:2014lha,Zhu:2014sya}. It's improper to
infer anything about SCC when superradiance occurs, since the perturbations
will be severely unstable even in the exterior of the black hole in
this case. Note that, the $l=0$ zero mode is trivial and should be
ignored in the limit of $q\rightarrow0$, so the subdominant mode
of $l=0$ should be considered like in \citep{Mo:2018nnu}. For a
nonzero $q$, we can confirm that SCC is respected, due to the nontrivial
lowest-lying of $l=0$.

\begin{figure}
\begin{centering}
\includegraphics[scale=0.45]{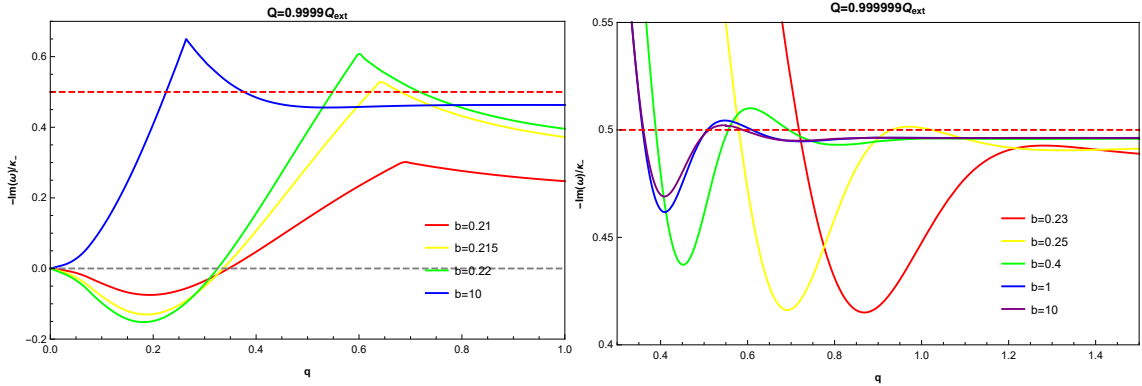}
\par\end{centering}
\caption{The lowest-lying QNMs $-\textrm{Im}(\omega)/\kappa_{-}$ of $l=0$
with varying $q$ for $\Lambda=0.14$ and various values of $b$.
Left: $Q=0.9999Q_{ext}$. Right: $Q=0.999999Q_{ext}$.}
\label{5}
\end{figure}

Since the lowest-lying QNMs of $l=0$ dominate $\beta$ for the charged
scalar field in a near-extremal black hole, we now focus on the $l=0$
mode for black holes further near the extremality. In FIG.\ref{5},
we plot the lowest-lying QNMs of $l=0$ with varying $q$ for $\Lambda=0.14$.
The left figure shows the case of $Q=0.9999Q_{ext}$. The blue line
represents the black hole of $b=10$, which is almost identical to
the case of RN-dS black hole. For $b=10$, $0.22$ and $0.215$, we
can find a narrow scalar charge window, where SCC is violated. Interestingly,
as $b$ decreases, the violation regime decreases, and for $b=0.21$,
the violation regime totally disappears. What's more, in the right
panel of FIG.\ref{5}, with $Q=0.999999Q_{ext}$, we can find wiggles
near $\beta=\frac{1}{2}$ in the cases of $b=10$, $1$, $0.4$ and
$0.25$, while in the case of $b=0.23$ the wiggle disappears. The
presence of the wiggles has been discussed in detail in the RN-dS
black holes in \citep{Dias:2018ufh}. We also find that the wiggle
shifts towards the direction of $q$ increasing as $b$ decreases.

\section{Conclusion}

In this paper, we investigate the validity of SCC for a massless scalar
field in a logarithmic-dS black hole. We first make a brief discussion
on the logarithmic-dS black holes and give the allowed region in which
the Cauchy horizon exists in the Section \ref{sec:Logarithmic-dS-Black-Hole}.
After that we make some preparations for the calculation of QNMs in
Section \ref{sec:Quasinormal-Mode}. Finally, we present the numerical
results of neutral scalar fields and charged scalar fields in Section
\ref{sec:Numerical-Results}.

It is noteworthy that when $b$ goes to infinity, the logarithmic-dS
black hole will go back to the RN-dS black hole as expected. Therefore
the behavior of SCC in a logarithmic-dS black hole is similar to that
of a RN-dS black hole when $b$ is big enough. When the NLED effect
increases, however, we find some interesting behaviors of SCC which
are different from RN-dS black hole. Through the analysis of the numerical
results, we found that the NLED effect can to some extent rescue SCC
for a near-extremal logarithmic-dS black hole. The specific impact
of NLED effect on SCC is as follows.

For a massless neutral scalar field: firstly, as NLED effect increases,
the minimal $Q/Q_{ext}$ for which the violation of SCC emerges goes
to $1$; secondly, given a fixed $Q/Q_{ext}$, the NLED effect can
always rescue SCC as long as the parameter $b$ goes to $b_{Q/Q_{ext}}$.

For a massless charged scalar field: firstly, the NLED effect can
lead to the appearance of superradiance; secondly, the NLED effect
can eliminate the narrow scalar charge window where SCC is violated;
thirdly, the NLED effect can eliminate the wiggles near $\beta=\frac{1}{2}$.

In general, no matter for the massless neutral scalar field or the
massless charged scalar field, the NLED effect is able to rescue SCC.
We find that the NLED effect of logarithmic-dS black holes can lead
to the shifting of the wiggles which is not obsevered in the Born-Infeld
case. Since the two NLED effects show great difference only when $b$
tends to $0$, it's reasonable that the two effects are similar with
$b>b_{min}$.
\begin{acknowledgments}
We are deeply grateful to Peng Wang and Bo Ning for their helpful
discussions and suggestions. This work is supported in partial by
NSFC (Grant No. 11505119, 11005016, 11875196 and 11375121).
\end{acknowledgments}

\bibliographystyle{unsrturl}
\bibliography{SCC}

\end{document}